\documentclass[preprint,showpacs,preprintnumbers,amsmath,amssymb]{revtex4-1}

\usepackage{graphicx}
\usepackage{dcolumn}
\usepackage{bm}
\usepackage{epsfig}

\begin{document}
\title{On physical insignificance of null naked singularities 
}
\author{ Shrirang S. Deshingkar}
\email[]{ssdeshingkar_at_gmail_dot_com}
\affiliation{
35, Ruikar Colony,
Kolhapur, 416005
India
}
\begin{abstract}
In this work we study collapse of a general matter in a most general
spacetime i.e., a spacetime with any matter and without (assuming)
any symmetry. 
We show that the energy is completely trapped 
inside the null singularity and therefore they cannot be experimentally observed.
This most general result implies,  there is  no physical significance of the null 
naked singularities irrespective of their existence. This conclusion strongly supports the essence of 
cosmic censorship hypothesis.

\end{abstract}

\pacs{0.42.Dw, 0.42.Jb}

\maketitle

\section{Introduction}
\label{s-intro}

Black holes are one of the most well studied objects in general relativity. 
When combined with theory, several astronomical observations exhibit 
strong evidence for existence of stellar mass (few times solar mass
$M_{\odot}$)~\cite{Casares:2006vx} 
as well as supermassive  (mass $10^5-10^9 M_{\odot}$)~\cite{2005SSRv..116..523F}
black holes.
The evidence of stellar mass black holes typically comes from observations of 
binary system, while by recent observations the supermassive black holes 
are expected to exist at the center of almost every galaxy.
In recent times there are some observational evidences~\cite{2009Natur.460...73F} 
for intermediate  mass black holes.
Existence of primordial black 
holes~\cite{Carr:2005zd} in the early universe is also speculated. Most of the primordial black 
holes are expected to have very small mass. 
Since the black holes have a trapped surface, singularity theorems~\cite{HawkEllis} guarantee 
a singularity inside their horizon, although, they are silent about their structural details. 
Singularities also form without or before formation 
of a trapped surface or an event horizon. If such a singularity forms, then
non-spacelike geodesics come out of it and in principle the singularity can be 
visible to an outside i.e., nonsingular observer; therefore, it is called a 
naked singularity. There are speculations~\cite{Witten:1992fp} 
that gamma ray bursts originate from such naked singularities.

We cannot foretell laws of physics at a singularity, hence existence 
of a naked singularity can lead to breakdown of predictability. Normally,
we do not want such a situation to arise in nature. With this in mind 
Penrose~\cite{Pen1} proposed the Cosmic Censorship Hypothesis (CCH). The CCH  
asserts  that,
naked singularities should not/do not form from collapse of a reasonable
matter field when we start with a generic nonsingular initial data. 
There
are many investigations supporting the CCH as well as many examples of existence
of naked singularities~\cite{HaradaICgc,EardlySmarr, ChirstoTBL, NewmanTBL, 
PsjTP,  SanjTBL1,  SarayGen, Dwivedi:1998ts,
PsjPrd93, ShrirLum1, ShrirTime, NolanNrng,  
Ori:1987hg, Ori:1989ps, HaradaPfluid, HaradaMena, 
MagliTan2, TangSanjMagGov, BarveTan, SanjCluster, PsjMahaGo, 
HaradaTimeFam, 
Lake, DwiPsjVaidya, 
MagliExact, SanjNeg,
GhoshDadhich2,   
PsjCmp94, ChrisScalar}.
Normally,  these investigations put several 
restrictions on the model. Almost all 
these studies assume a very specific and very restrictive form of matter.
They also assume only a certain kind of metric, typically, either with spherical or axial symmetry.
Therefore, they may not satisfy the genericity of initial
data or reasonable matter field criteria.
Hence, they do not prove or disprove the CCH.  The main difficulty  arises as
in general we cannot establish the relationship of initial data to the
formation and structure of the singularity even after assuming lot of symmetry and taking
a very simplistic equation for matter.
Therefore, proving or disproving the CCH remains one of the most important open
problems in classical general relativity.

In such a situation we need to investigate, whether some other effects safeguard the 
physics.  
It may happen that, even if naked singularities exist geometrically, 
they have no physical or observational consequences. i.e. they cannot
affect the world outside them and other physical effects make them benign.
Naked singularities
and the CCH get lot of attention and is important as general relativity is 
a physical theory which describes nature. 
Unpredictability is a serious issue for a physical theory and needs to be fixed.
Therefore, it is important to find out whether a naked singularity can affect the
physics outside it, i.e., whether a nonsingular observer can distinguish it observationally.
With this in mind we studied the spherically symmetric dust
collapse model~\cite{CanWe1} and calculated the redshift and luminosity of
light rays coming out of the singularity. Subsequently, we also studied
extremely wide class  of (any type II matter) spherically symmetric collapse models for the same
purpose. In both the cases we showed that at the most along one singular
null geodesic coming out of a null (naked) singularity the redshift 
is finite while redshift diverges along 
all other outgoing singular null geodesics.  Hence, we concluded that 
the null naked 
singularities will not be physically troublesome, as no energy can come out of
them.   
 
Here, we generalize the proof for any null singularity.
In this work, we consider a completely general spacetime metric without
 any symmetries. Apart from weak energy condition~\cite{HawkEllis}, no restrictions on form of matter are
assumed.
We show that if a null singularity forms in the collapse,
then at the most for one singular null geodesic the redshift is finite while
it is infinite for all other (infinite family) of singular geodesics.
Hence no energy or information can come out
of a null naked singularity. We need a wavepacket to carry energy, thereupon,
redshift should be finite for a finite (though it can be very small) 
duration to get out energy from the singularity. As no energy can come out
of the null singularity, it cannot affect the physics outside and there
should not be any danger of breakdown of predictability. This means, 
physically the null naked singularities are not important or they are not dangerous.

We also show that the redshift is always finite for null geodesics coming out
of a timelike naked singularity. This  means that, in
principle  such a singularity can be observed and they can be 
more problematic. However, as such we expect them to be very rare~\cite{CanWe1}.
           
\section{methodology}
\label{s-methodology}

        We use the characteristic or null cone formulation. 
The spacetime is foliated using a family of outgoing null
hypersurfaces. We use the Bondi-Sachs~\cite{Bondi:1962px,Sachs:1962wk} coordinates. These coordinates
are commonly used in characteristic approach to numerical 
relativity ~\cite{Lehner:2001wq, Bishop:1999yg, Bishop:1997ik,Siebel:2003sp}. 
The null hypersurfaces are labelled by  coordinate $u$; $r$ is the
area radius coordinate and  the angular coordinates $x^A$ $(A=2,3)$
label the outgoing null rays. The metric in Bondi-Sachs 
coordinates~\cite{Bondi:1962px, Sachs:1962wk} is written as 
\begin{eqnarray}
   ds^2  =  -\left (  \frac{V}{r} e^{2\beta}  -r^2h_{AB} U^A U^B \right ) du^2
\nonumber \\
      - 2e^{2\beta} du dr -2r^2 h_{AB}U^B du dx^A
       +  r^2h_{AB}dx^A dx^B,
\label{eq:BSmetric}
\end{eqnarray}
where $h^{AB}h_{BC}=\delta^A_C$ and
$det(h_{AB})=det(q_{AB})$, with $q_{AB}$ a unit sphere metric.
All the metric variables, i.e., $\beta$, $V$, $U^A$ and $h_{AB}$  are
functions of $(u,r, x^A)$. 
This metric in general does not have any
symmetries and any spacetime can be written in this form.
The field $\beta$ gives expansion of light rays, $V$ is akin to the Newtonian
potential, $h_{AB}$ gives the conformal geometry of $r=\rm constant$, 
$u = \rm constant$ surface
(or the deviation from spherical symmetry), 
$U^A$   represents the shift vector.
$e^{2\beta} V/r$ is analogous to square of the lapse function.

         The Einstein equations are decomposed as constraint equations and
evolution equations.  The constraint equations are basically hypersurface 
equations. For our purpose here, 
we need only a couple of 
hypersurface equations, which are written 
as~\cite{Lehner:2001wq, Bishop:1999yg, Bishop:1997ik}
\begin{eqnarray}
r R_{rr} = & 4 \beta_{,r} -  \frac{1}{4}rh^{AC}h^{BD}h_{AB,r}h_{CD,r} \: ,
\label{eq:R_rr}
\\
r^2 g^{AB}R_{AB}   = &- 2e^{-2\beta}V_{,r} + {\cal R} - 2 D^{A} D_{A} \beta
\nonumber\\
&-2 D^{A}\beta D_{A}\beta 
+ r^{-2} e^{-2\beta} D_{A}(r^4U^A)_{,r} \nonumber\\
&-\frac{1}{2}r^4e^{-4\beta}h_{AB}U^A_{,r}U^B_{,r}, 
~
\label{eq:gABRab}
\end{eqnarray}
where  $D_A$ is the covariant derivative and ${\cal R}$ the curvature
scalar of the 2-metric $h_{AB}$. We do not need the evolution equations
for our purpose.

           At the regular center various metric quantities have to obey some 
regularity conditions. They are given as, 
$U^A = O(r), h_{AB} = q_{AB} + O(r^2),  V = r + O(r^3)$ and $\beta = O(r^4)$.
The form of $h_{AB}$ near the center also tell us that, at the center 
we have ${\cal R} = 2$.
As we are using the outgoing null coordinates, $du = dx^A = 0$ represent
the outgoing null geodesics. We will call these Outgoing Radial Null 
Geodesics (ORNGs). The geodesic tangent vector $K^r$
is given as
\begin{equation}
K^r = \frac{dr}{d\lambda}= C_1 e^{-2\beta},
\label{eq:Kr}
\end{equation}
where $\lambda$ is the affine parameter and $C_1$ is a constant of 
integration.

Let ${u^a}_{(s)}$ and ${u^a}_{(o)}$ be the four-velocities of the
source and the
observer and let $E_1$ and $E_2$ be two events connecting the source
and the observer through the ORNG. The redshift $z$ is given 
by~\cite{1990rcm..book.....D,1991grra.book.....S}
\begin{equation}
1+z = \frac{[K_a {u^a}_{(s)}]_{E_1}} { [K_a {u^a}_{(o)}]_{E_2} },
\end{equation}
where the numerator and denominator are evaluated at events $E_1$ and $E_2$,
at the source and observer respectively, with
$ 
{u^a}_{(s)}= -(  e^{2\beta}  {V}/{r} -r^2h_{AB} U^A U^B)^{-1/2}_{(s)}
\delta^a_u 
$
and
$
{u^a}_{(o)}=  -(  e^{2\beta} {V}/{r}   -r^2h_{AB} U^A U^B )^{-1/2}_{(o)}
\delta^a_u $.
Taking the source as the naked singularity at $r=0$ or $r = \rm constant$ 
we get 
\begin{equation}
1+z =  \frac { \left (  \frac{V}{r} e^{2\beta} 
-r^2h_{AB} U^A U^B   \right )^{1/2}_{(o)}} 
{  \left (  \frac{V}{r} e^{2\beta} 
-r^2h_{AB} U^A U^B   \right )^{1/2}_{(s)} }   .
\label{eq:redshift2}
\end{equation}

We assume that a singularity forms in the gravitational collapse.
We can choose our coordinates such that a point where the singularity
forms is at $r= \rm constant $. 
Now if the singularity is null singularity,
then after it forms
$ r= \rm constant $ should be null, i.e. we should 
have $ dr/du = 0 $ for the ingoing
null geodesic, while outgoing null geodesics will be 
represented by $du=0$ by our coordinate choice.  This gives
\begin{equation}
\frac{dr}{du} = \left (\frac {V} {r} -r^2h_{AB}U^AU^B e^{-2\beta}\right) = 0.
\label{eq:NullSing}
\end{equation}
This with Eq.~(\ref{eq:redshift2}) tell us that the redshift diverges
for geodesics coming out of a null singularity. 
Our argument may not apply to the first point of 
singularity formation as it is a boundary point and there the redshift 
could be finite or infinite.  But that is just an instant and finite (non-zero)
amount of energy cannot come out of it.

We can choose coordinates suitable for our purpose, e.g., we
can choose them such that the singularity forms at $r=0$. 
In spherically symmetric case we can choose $h_{AB}=q_{AB}$
and $U^A = 0$ and equations~(\ref{eq:R_rr},\ref{eq:gABRab}) simplify
a lot. Different choices for $U^A$ will lead to different sets of null 
geodesics along $u=constant$, $r=constant$. 
In spherically symmetry a shell focusing naked singularity can
form only at $r=0$~\cite{CanWe1} when weak energy condition~\cite{HawkEllis} 
holds.

         If the $dr = dx^A = 0$ curve is timelike, then 
from the metric (Eq.~(\ref{eq:BSmetric})), 
$(e^{2\beta} {V}/ {r} -r^2h_{AB}U^AU^B ) $ is finite and from 
Eq.~(\ref{eq:redshift2}) we get that the redshift is always finite.
Please note that if $e^{2\beta}$ diverges, then weak energy condition does not
hold from Eq.~(\ref{eq:R_rr}).
If the singularity 
is spacelike, then it is always covered and the question of redshift 
of  radiation coming out of the singularity does not arise.

We have checked doing the coordinate transformations from the commonly used
coordinates for various spherically symmetric models to the
retarded Bondi-Sachs  coordinates near the central singularity, namely, for
the cases of
dust~\cite{ShrirTime, ShrirLum1}, general matter~\cite{CanWe1} and also 
Vaidya radiation collapse~\cite{DwiPsjVaidya}.
Those calculations also confirm that our conclusions are
correct. 

\begin{figure}[t]
\includegraphics[height=5cm]{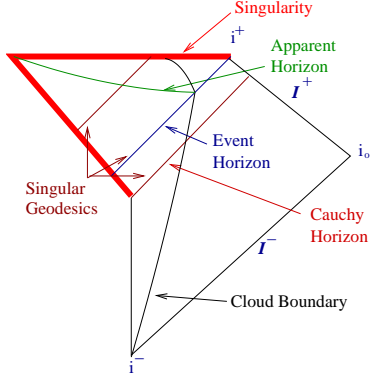}
\caption{Penrose diagram for a globally naked null singularity}
\label{fig:PenGlobNaked}
\end{figure}

As such we have not used $r=\rm constant$ is a singularity to get the
redshift and in general the curvature or the Einstein equations will tell us if it is
a singularity. In most cases some components of curvature diverge at singularity.
We basically derive the redshift using only the geometry of the
spacetime and do not explicitly need the Einstein equations. Hence the 
results are extremely general. Just the proper time of a (singular) source
and an observer tells us that the redshift for any null (singularity) surface
will diverge. Basically if $\Phi$=0 is ekonal of the wave, then the frequency
in geometrical optics approximation is~\cite{1990rcm..book.....D,1991grra.book.....S}
\begin{equation}
\nu= \frac{d\Phi}{dT}, 
\label{eq:frequency}
\end{equation}
where $T$ is the proper time. Redshift is ratio of the frequency 
at source to the frequency at observer. For the null source proper time is 
zero i.e. $dT=0$ and so redshift  diverges for any observer which is not 
at the source. 
In figure~\ref{fig:PenGlobNaked} we have shown a Penrose 
diagram for a globally naked null singularity. It is clear from the Penrose 
diagram that no ingoing timelike geodesic can reach the null singularity. 
Any source which is at the singularity for finite time has to be null. 
That means the redshift along any rays coming out of the null singularity 
will be infinite.
For a timelike source, the proper time is finite i.e., nonzero;
therefore, redshift is always finite, though it can be very large.

        The observed intensity $I_p$ of a point source~\cite{DwiKan} is
\begin{equation}
I_p= \frac{{P_0}}{{A_0 (1+z)^2}},
\label{eq:luminosity}
\end{equation}
where $P_0$ is the power radiated by the source into the solid angle
$\delta\Omega$, and $A_0$ is the area sustained by the rays at the observer.
The redshift factor $(1+z)^2$ appears because the power radiated is not
the same as power received by the observer; 
there is a effect of time dilation as well as frequency change. 
As the redshift diverges
along  geodesics coming out of a null (naked) singularity and $A_0$ is finite,
classically the observed luminosity of a null naked singularity is zero. It means that no 
energy can reach an outside (nonsingular) observer from any null singularity.
        In essence, we need the redshift to be finite for a finite period
to carry any energy outside from a source. For null singularities it is finite 
at the most for an instant i.e., along the first point/ray coming out of the null
naked singularity.  Hence no energy can reach outside
from a null naked singularity; it cannot affect the physics
outside and therefore, there will be no breakdown of predictability. 
A timelike singularity can  in principle be visible to an outside 
observer as the redshift is always finite  for it.

Our result about redshift and luminosity is not very 
surprising and is also expected from special
relativity. In the special theory of relativity, as the source 
velocity approaches velocity
of light the redshift of the emitted radiation starts diverging and the emission
cone becomes narrower and narrower. When the source is moving at the speed of 
light all the radiation is directed only in that source's direction and for other rays
going out from the source redshift is infinite. In our case the source is  null
(singularity) and so for any observer not sitting on the source (singularity) the
redshift of the rays coming to it from the null source is infinite and energy 
reaching it from the null source is zero. That means our general relativistic 
result is expected from special relativity.

\section{Summary}
\label{s-summary}
         In summary, in this work, writing the metric in the 
Bondi-Sachs form  
we have shown that no energy
can come out of any null naked singularity. 
The advantage of using null formulation is that it is designed to study
the null geodesics which we want to explore. Hence no extra work is needed
to get out the quantities of our interest.
We have not imposed any 
symmetries on the spacetime nor we have assumed any specific form of
matter. 
Essentially the assumption
that the singularity which forms is null and just the geometry of spacetime
is enough for us to reach this conclusion. 
          We also showed that a timelike singularity is in principle likely to be visible 
to an outside/nonsingular observer as redshift is always finite for rays
coming out from it. However, we  expect formation of a timelike
singularity to be rare when energy conditions hold, as it seems~\cite{CanWe1}
creation of such a singularity needs
fine balance of factors;  we require the collapse to stop as soon as
the singularity is formed. That indicates
they will be non-generic. The known examples of naked singularities are
also mostly null, or else, they are formed in a very unusual matter fields and need
lot of fine tuning.

 We have also given the fundamental principle/logic 
which drive the result. 
As we have explained above,
the result is essentially very similar to the special relativistic result for
a null (or timelike) source. For the null singularity (surface) the
redshift basically diverges as the proper time goes to zero on null surface.
So if a ray has finite frequency outside the singularity (surface) and it 
originates at the singularity, then it has to have infinite frequency at 
the singularity. Similarly for timelike singularity as the proper time is 
always finite (nonzero) at the singularity so the redshift is finite, though
it can be very large.

Our result is valid for any form of matter as it is purely based on
the geometry at the source i.e., geometry at the singularity. 
That means our result(s) are most general and should be valid in any theory 
of gravity  as well as, in higher dimensions. 
One can generalize these
results easily along the timelike geodesics. 
The results basically means that, though the null singularity is 
geometrically naked (i.e., null geodesics can come
out of it) essentially physically it is not visible (naked), as no
energy can come out of it due to infinite redshift.
That implies we cannot get 
any information from the null naked singularity and it will not have 
any undesirable physical effect outside. Therefore, null naked singularity
cannot cause breakdown of predictability and they have no special physical 
significance; i.e.  the fact that non-spacelike geodesics come out of
null naked singularity does not have any significance. Our conclusion is
extremely general and is valid for all possible spacetimes. This strongly
supports/preserves the essence of comic censorship for null singularities.
However, we need to study the timelike naked singularities
in more detail for this purpose.

I would liked to thank IUCAA, Pune and CTP, Jamia Millia Islamia, Delhi for hospitality
where early part of this work was done.

\bibliography{gccf}
\end{document}